\documentclass[showpacs,aps,pra,twocolumn,amsmath,amssymb,superscriptaddress]{revtex4}
\usepackage{amsmath}
\usepackage{graphicx}
\usepackage{dcolumn}
\usepackage{color}
\usepackage{bm}

\begin{document}

\title{Spontaneous synchronization and quantum correlation dynamics of open spin systems }

\author{G.~L.~Giorgi}
\affiliation{INRIM, Strada delle Cacce 91, I-10135 Torino, Italy }
\affiliation{AG Theoretische Quantenphysik, Theoretische Physik, Universit\"at des Saarlandes,
D-66123 Saarbr\"ucken, Germany}

\author{ F.~Plastina}
\affiliation{ Dipartimento di  Fisica, Universit\`a della
Calabria, 87036 Arcavacata di Rende (CS), Italy}
\affiliation{  INFN -
Gruppo collegato di Cosenza, Cosenza (Italy)}

\author{G.~Francica}
\affiliation{ Dipartimento di  Fisica, Universit\`a della
Calabria, 87036 Arcavacata di Rende (CS), Italy}

\author{R.~Zambrini}
\affiliation{ IFISC (UIB-CSIC), Instituto de
F\'{\i}sica Interdisciplinar y Sistemas Complejos, UIB Campus,
E-07122 Palma de Mallorca, Spain}


\pagenumbering{arabic}

\date{\today}

\begin{abstract}
We discuss the emergence of spontaneous synchronization for an
open spin-pair system interacting only via a common environment.
Under suitable conditions, and even in the presence of detuning
between the natural precession frequencies of the two spins, they
are shown to reach a long-lasting transient behavior where they
oscillate in phase. We explore the connection between the
{emergence} of such a behavior and the establishment of robust
quantum correlations between the two spins, analyzing differences
between dissipative and dephasing effects. In particular, in the
regime in which synchronization occurs, quantum correlations
are more robust for shorter synchronization times and this is
related to a separation between system decay rates.
\end{abstract}

\maketitle

\section{Introduction}
Synchronization is a paradigmatic phenomenon in complex systems,
characterized by a coherent dynamics of different oscillating
units \cite{Pik,Str}. Spontaneous synchronization generally arises
in spite of detuning of the natural frequencies  of component
subsystems, due to their weak interaction. After more than three
centuries from the first reported observation \cite{Huygens}, this
phenomenon has been identified in several physical, biological,
chemical and social systems \cite{Pik,Str,Arenas}. At the
microscopic level, mutual synchronization has been studied in
different devices, such as arrays of Josephson junctions
\cite{Wiesenfeld}, spin torque nano-oscillators \cite{Kaka},
and nanomechanical  \cite{Shim} and optomechanical oscillators
\cite{Marquardt2011,Milburn2011,Lipson2012,mari}. Most of these
implementations at micro- and nanoscale have focused on the
classical dynamics, while quantum fluctuations and correlations
have been analyzed in \cite{PRAsync,3HO,sync_net,mari}.

A full quantum approach has been recently reported  for forced synchronization
by Goychuk \textit{et al.} \cite{Hanggi} considering quantum stochastic
synchronization in  the spin-boson dynamics in the presence of a
driving signal modulated in time and reporting on the constructive
role of thermal noise. Furthermore, synchronization with driving
was considered by Zhirov and Shepelyansky, {who discussed} the
effect of a driven resonator in the cases of both one
\cite{zhirov1}  and two superconducting qubits \cite{zhirov2}.
These works explore the phenomena of forced synchronization,
{usually referred to as} entrainment, in the quantum regime where
the system synchronizes with the external driver {instead of
following its natural frequency}. The presence of driving out of
equilibrium does also favor quantum effects \cite{galve}.

On the other hand, {\it spontaneous} or mutual synchronization
between detuned coupled systems is the coherent dynamic phenomenon
of rhythm adjustment  without any external driver taming the
evolution. The emergence of spontaneous quantum synchronization
has been recently considered for dissipative harmonic oscillators
with two major breakthroughs: (i) the possibility to have
synchronization induced by dissipation in a linear system and (ii)
the  full quantumness of this phenomenon
(reported for vacuum fluctuations)
 \cite{PRAsync,3HO,sync_net}. Synchronous
dynamics has been reported during the relaxation process, in spite
of the diversity of the natural frequencies of a pair of
oscillators  \cite{PRAsync}, due to the occurrence of a slowly
decaying mode responsible for synchronization
 accompanied by robust and asymptotic quantum
correlations in the system \cite{PRAsync,3HO}.
Interestingly, if the oscillators experience losses in separate
baths, synchronization does not emerge, independently of the
strength of their coupling \cite{PRAsync}.
When more than two detuned oscillators are considered and
depending on the environment correlation length, synchronization
and robust quantum correlations can arise not only in a transient
but even $asymptotically$ \cite{3HO,sync_net}, associated with the
presence of some decoherence-free normal mode of the system, as
reported even for random networks \cite{sync_net}.

\begin{figure}
\includegraphics[width=8cm]{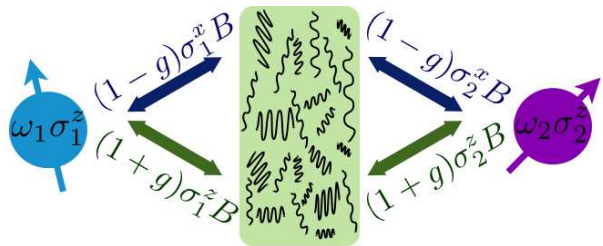}
\caption{(Color online) Schematics of the system. Two spins
interact with a common thermal bosonic  environment [and
specifically to the quantum and thermal fluctuations of the
environmental operator $B=\sum_k \gamma_k(a_k^\dag+a_k)$] through
the two different coupling mechanisms reported in Eq. (\ref{vs}).
The first coupling term, $ \propto (1+g)\sigma_i^z$, only induces
pure dephasing, while the second one, $\propto (1-g)\sigma_i^x$
($i=1,2$), is also a source of dissipation. \label{fig1}}
\end{figure}

A main open question is about the possibility to induce
synchronization in the presence of a different kind of coupling to the
environment, giving rise to dephasing rather than dissipation. In
this paper, we tackle this question by discussing the dynamics of
two precessing spins [see Fig. (\ref{fig1})] detuned from each
other and experiencing decoherence due to the coupling with an
environment, with the aim of assessing in such a framework the key
mechanism responsible for quantum synchronization. As for the case
of quantum oscillators \cite{sync_net}, we will show that the
form of the coupling with the bath is a crucial ingredient for a
synchronous dynamics to emerge between detuned quantum subsystems
relaxing towards equilibrium.
In order to establish their distinctive roles,  both a dissipative
and a purely dephasing spin dynamics will be studied, and, through
a sensible parametrization of the system-bath interaction, a
continuous transition between these two extreme cases will be
considered.

Another question that naturally arises about synchronization in
the quantum regime is whether it is related to the appearance of
entanglement or of more general quantum correlations, measured,
e.g., by quantum discord whose dynamics has been studied
extensively for a detuned spin pair in a common environment (see,
for instance, Refs. \cite{francica,benfanchini}).

Comparing with previous works on synchronization in the quantum
regime
\cite{PRAsync,3HO,sync_net,mari,Marquardt2011,Milburn2011,Lipson2012,liu},
an important difference is that  we are going to consider spins
that are not directly {coupled to each other},  so that
spontaneous synchronization as well as quantum correlations only
arise due to the indirect, bath-mediated coupling. Mutual
synchronization in  nonresonant spin systems was also studied by
Orth \textit{et al.} in Ref. \cite{lehur}, analyzing  the case of
two spins coupled to each other via an Ising-like term and
strongly interacting with a common bath, which induces a
substantial renormalization of the coupling strength. Under these
conditions, a few correlated and synchronous oscillations are
observed before dissipation prevails. Here, instead, we are going
to consider the weak coupling regime between spins and bath, to
show that a long-time, robust, synchronous dynamics occurs, even
in the absence of a direct coupling between the subsystems.

\section{Description of the Model}\label{model}

We consider two noninteracting qubits with different precession
frequencies $\omega_1^\prime$ and $\omega_2^\prime$ and coupled to a common thermal bath. By employing
units in which $\hbar =1$, the system Hamiltonian can be written as
\begin{equation}
 H_S=\omega_1 \sigma_1^z+\omega_2 \sigma_2^z,
\end{equation}
where $\omega_1=\omega_1^\prime/2$ and
$\omega_2=\omega_2^\prime/2$. The common thermal environment is
modeled as a set of independent harmonic oscillators, $H_{B}=
\sum_k \Omega_k a_k^\dag a_k$, taken in its thermal state
$\rho_B=\exp[-\beta H_B]/{\rm Tr}\{\exp[-\beta H_B]\}$, and the
system-bath interaction term has the form
\begin{equation}
 H_{SB}=  V_{S}\sum_k \gamma_k(a_k^\dag+a_k).
\end{equation}
A sketch of the model is given in Fig. \ref{fig1}. Generically, in
a spin-boson problem, the environmental fluctuations are coupled
both longitudinally and transversally to the spin, e.g., via
interaction terms proportional to $\sigma^z$ and $\sigma^x$,
respectively (see \cite{weiss_book)}. As we shall see throughout
the paper, a longitudinal coupling, in which the bath only induces
dephasing in the system without any dissipation, plays a special
role. On the other hand, synchronization is found to occur
essentially when the transversal coupling (inducing relaxation)
overcomes the longitudinal one. To discuss these issues in the
simplest possible way, we model the system-environment coupling
$V_S$ as
\begin{equation}
 V_{S}=(1+g)(\sigma_1^z+ \sigma_2^z)+ (1-g)(\sigma_1^x+
 \sigma_2^x)\label{vs},
\end{equation}
where we have introduced an anisotropy coefficient $g \in [-1,1]$.

Notice that, in the limit of $g=1$, the system Hamiltonian $H_S$
commutes with the total Hamiltonian $H=H_S+ H_{B}+ H_{SB}$ and no
energy exchange between system and environment can take place,
while, for $g \neq 1$, the dynamics of the spin pair always
includes some degree of relaxation. The two relevant parameters of
the system are the coefficient $g$ and the detuning
$\Delta=\omega_2-\omega_1$. Henceforth, we shall take $\omega_1$
as the scale of energy and inverse time, and therefore, from
now on, all of the frequencies are evaluated in units of
$\omega_1$.

Through a rotation, the Hamiltonian model introduced here can be
mapped into the more common one describing two Josephson qubits in
a noisy environment \cite{yuri,storcz}. In fact,  $H$ is unitarily
equivalent to a set of possible realizations of $H^{'} = H_S^{'} +
H_{SB}^{'}$, where the system Hamiltonian and the system-bath
interaction Hamiltonian would read, respectively $
H_S^{'}=\Delta_1 \sigma_1^x+\Delta_2 \sigma_2^x+\epsilon_1
\sigma_1^z+\epsilon_2 \sigma_2^z$ and $
H_{SB}^{'}=\sqrt{2(1+g^2)}(\sigma_1^z+\sigma_2^z)\sum_k
\gamma_k(a_k^\dag+a_k)$, with the {constraint}
$\Delta_1/\epsilon_1=\Delta_2/\epsilon_2=(g+1)/(g-1)$.

The time evolution of the reduced density matrix of the two spins
can be calculated using the Bloch-Redfield master equation
approach \cite{weiss_book}. Up to the second order in the
system-bath coupling and in the Markov approximation, we find the
following set of equations of motion for the matrix elements of
the reduced density matrix in the basis of the eigenstates of
$H_S$:
\begin{equation}
\dot \rho_{ab}=-i\omega_{ab} \rho_{ab}-\sum_{mn}R_{abmn}\rho_{mn}, \label{roab}
\end{equation}
where $\omega_{ab} = E_a - E_b$, and where $E_i$ are the eigenvalues of the unperturbed two-qubit
Hamiltonian. The elements of the Redfield tensor are given by
\begin{eqnarray}
R_{abmn}&=&\delta_{bn}\sum_r S_{ar} S_{rm} \Gamma^+(\omega_{rm})
-S_{am}S_{nb} \Gamma^+(\omega_{am})
\nonumber\\&+&\delta_{am}\sum_r S_{nr} S_{rb} \Gamma^-(\omega_{nr})-S_{am}S_{nb} \Gamma^-(\omega_{nb})\nonumber\\
\end{eqnarray}
where $S_{ij}=\langle i |V_S| j \rangle$. Introducing the bath
density  of states $J(\omega)= \sum_k \gamma_k^2
\delta(\omega-\Omega_k)$, the coefficients $\Gamma^{\pm}$ read
\begin{eqnarray}
\Gamma^\pm(x)&=&\frac{\pi}{8}   [J(x)+J(-x)]
\left(\coth\frac{\beta x}{2}\mp
1\right)\nonumber\\&+&\frac{i}{4}{\rm
P}\int\frac{J(\omega)}{\omega^2-x^2}\left( x \coth\frac{
\beta\omega}{2} \mp \omega\right)d\omega,
\end{eqnarray}
where ${\rm P}$ denotes the Cauchy's principal value and where
$\beta=1/ k_B T$ is the inverse temperature of the bath. We shall
assume an Ohmic environment with a Lorentz-Drude cut-off function,
whose spectral density is
\begin{equation}\label{JOhm}
 J(\omega)=\gamma \omega \frac{\omega_c^2}{\omega_c^2+\omega^2}.
\end{equation}
The cut-off $\omega_c$ is bounded to ensure the validity of the
Markovian approximation, $\omega_c\gg\omega_i$, and also for the
Bloch-Redfield master equation to give a correct estimation of the
renormalizing effects of the bath, $\gamma \omega_c\ll\omega_i$
($i=1,2$).

It is important to remark here that, in general, the
Bloch-Redfield second order master equation
 is known to provide neither a completely positive nor a positive map
(see, for instance, \cite{benatti} for a review). However, usually,
if the system-bath coupling is weak enough, as compared to the energy scales of the system, positivity is not violated.
 For all the results  presented hereafter,
we have numerically checked the positivity of the reduced density
matrix of the system from the beginning of the evolution (where
system and bath are assumed to be uncoupled) until the system
reaches or is very close to reaching a final state that can be
stationary or oscillating depending on the existence of noise-free
channels.

\subsection{Decoherence free evolution}
In the following section, we will show the occurrence of a {\it
significant} time window in which the local observables of the two
spins show synchronized oscillations before equilibration takes
place. We will require this transient regime to be robust enough
(e.g., to last for a very long time), and this will be linked to
the appearance of slowly decaying solution of the Bloch-Redfield
master equation (\ref{roab}).

In this respect, it is well known that special instances exist in
which equilibration does not fully take place because some state
of the system happens to be robust against decoherence
\cite{massimo,decohfree}, and that this can lead to asymptotic
quantum correlations \cite{maniscalco}. Indeed, despite the
presence of a thermal environment, there are cases where a
noiseless evolution can be observed, provided that the full
Hamiltonian possesses some special symmetry and that the initial
state of the system belongs to a given {\it decoherence free}
subspace. Such subspaces are found to exist for the cases of 
(a) identical spins ($\omega_1=\omega_2$), or (b) purely
dephasing dynamics ($g=1$). In the second case, in particular, the
model Hamiltonian becomes exactly solvable and its solution
describes the only instance in which the system is not totally
dissipative.

Strictly speaking, from the point of view of synchronization, the
first of these cases, (a), is trivial  (as the precession
frequencies are already equal), while the second, (b), is irrelevant
(as discussed in Sec. \ref{sec3}). However, their analysis will
appear to be crucial to understand the emergence of
synchronization (and its absence in some regime), and we briefly
recall it here.

\subsubsection{Identical spins} By assuming $\Delta=0$,
irrespective of the value of $g$, the maximally entangled state $
|\psi^-\rangle=(|\uparrow,\downarrow\rangle-|\downarrow,\uparrow\rangle)/\sqrt{2}
$ belongs to the kernel of  both $H_S$ and $V_S$ \cite{nota}.
Then, its evolution turns out to be decoherence-free. The special
role played by $|\psi^-\rangle$ gives rise to important
consequences to the long time behavior of the system, as for any
initial condition not orthogonal to this state the  system will
never reach a steady state.
 To make clear the importance of
initial conditions chosen in the figures presented hereinafter, we notice that
 a family of states that do reach a stationary condition
  at very long times is
given by symmetric factorized states, that is $(\cos \theta
|\uparrow\rangle + e^{i\phi}\sin \theta |\downarrow\rangle)\otimes
(\cos \theta |\uparrow\rangle + e^{i\phi}\sin \theta
|\downarrow\rangle)$, while asymmetric factorized states are in general not orthogonal to
 $|\psi^-\rangle$.

\subsubsection{Pure dephasing}
 For $g=1$, and independently of the
detuning, the bath can only induce dephasing, without any
dissipation, since $[H_S,V_S]=0$ \cite{massimo,quiroga}. The
dynamics of the reduced density matrix of the system can be
calculated exactly using, for instance, the coherent state method
introduced in Refs. \cite{massimo,privman}. Labeling with $E_i$
($\lambda_i$) the eigenvalues of $H_S(V_S)$, the density matrix
elements, written in the basis of the eigenstates of both $H_S$
and $V_S$, evolve according to
 \begin{equation}
 \rho_{ab}(t)=\rho_{ab}(0)e^{-i\omega_{ab}t}e^{-\sum_k\frac{|\gamma_k|^2}{\Omega_k^2}P_{ab,k}},
\end{equation}
with
\begin{eqnarray}
 P_{ab,k}&=& i(\lambda_a^2-\lambda_b^2)\sin\Omega_k t\nonumber\\&+&2(\lambda_a-\lambda_b)^2\sin^2\frac{\Omega_k t}{2}\coth\frac{\beta \Omega_k }{2}.
\end{eqnarray}
In the continuum limit we have
\begin{equation}
 \rho_{ab}(t)=\rho_{ab}(0)e^{-i\omega_{ab}t}e^{-(\Gamma_{ab}+i L_{ab})},\label{romn}
\end{equation}
where
\begin{equation}
 \Gamma_{ab}=2(\lambda_a-\lambda_b)^2\int d\omega J(\omega)\omega^{-2}\sin^2\frac{\omega t}{2}\coth\frac{\beta \omega }{2}
\end{equation}
and where the Lamb shift is
\begin{equation}
 L_{ab}=(\lambda_a^2-\lambda_b^2)\int d\omega J(\omega)\omega^{-2}\sin\omega t.
\end{equation}
The bath effect is entirely contained in
$\exp[-(\Gamma_{ab}+i L_{ab})]$. While $\Gamma_{ba}$ fixes the
dephasing rate, $L_{ab}$ introduces a shift in the oscillatory
dynamics.
In the weak coupling limit, $L$ is usually negligible and
$\Gamma(t)$ is a linear function of time. Notice that the detuning
$\Delta$ does not play any role in the decoherence process.

The existence of a common basis for $H_S$ and $V_S$  in the pure
dephasing limit implies that the four basis states
$|1\rangle\equiv|\uparrow\uparrow\rangle,|2\rangle\equiv
|\uparrow\downarrow\rangle,|3\rangle\equiv|\downarrow\uparrow\rangle,
|4\rangle\equiv|\downarrow\downarrow\rangle $ would evolve
without experiencing any kind of decoherence. Furthermore, since
$V_S|\uparrow\downarrow\rangle=V_S|\downarrow\uparrow\rangle=0$,
then, $ L_{23}=L_{32}=\Gamma_{23}=\Gamma_{32}=0$, and the whole
subspace spanned by the states $|\uparrow\downarrow\rangle$ and
$|\downarrow\uparrow\rangle$ is decoherence-free; that is, only
the oscillations due to the free evolution $H_S$ are displayed for
such initial states. Among other consequences, a maximally
entangled state belonging to this subspace ($|\psi^+\rangle$ or
$\psi^-\rangle$) will remain maximally entangled forever: for
instance, by considering $\rho_+(0)=|\psi^+\rangle
\langle\psi^+|$, we have
\begin{eqnarray}
\rho_+(t)&=&\frac{1}{2} \big( |\uparrow,\downarrow\rangle
\langle \uparrow,\downarrow | +  |\downarrow,\uparrow\rangle
\langle \downarrow,\uparrow | \nonumber\\&+&e^{2 i \Delta t}
|\downarrow,\uparrow\rangle\langle \uparrow,\downarrow | +e^{-2 i
\Delta t}  |\uparrow,\downarrow\rangle\langle \downarrow,\uparrow
|  \big).
\end{eqnarray}

\section{Synchronization}\label{sec3}
In this section, in analogy with what {was} done for the case
of harmonic oscillator systems in Refs. \cite{PRAsync,3HO,sync_net}, we
quantitatively discuss the emergence of synchronization and study
the conditions under which the dynamics of two spins with
different precession frequencies can exhibit a long-lasting
transient regime of phase-locked spin oscillations.

A full characterization of the {\it local} precessing dynamical
behavior of the two spins can be obtained by analyzing  the
evolution of the average values of their respective Pauli
operators. In the absence of noise, the values of  ${\langle
\sigma_1^x (t)\rangle}=2 {\rm Re}\{\rho_{13}(t)+\rho_{24}(t)\}$
and   $ {\langle \sigma_2^x (t)\rangle}=2  {\rm
Re}\{\rho_{12}(t)+\rho_{34}(t)\}$ would oscillate in time with
their respective frequencies $2\omega_1$ and $2\omega_2$ (the same
argument could be applied to $\langle \sigma_1^y\rangle$ and
$\langle \sigma_2^y\rangle$ or to any combination in the $x-y$
plane as well. On the other hand,  $\langle\sigma_1^z\rangle$ and
$\langle\sigma_2^z\rangle$ would be constants of motion. Then, any
couple of local spin operators lying in the $x-y$ plane is a good
candidate to test whether or not the bath is able to induce 
synchronous oscillations. For concreteness, in the following we
will consider $\langle \sigma_1^x\rangle$ and $\langle
\sigma_2^x\rangle$ and study the conditions under which these
variables synchronize. Actually, as we shall see, the form of the
interaction Hamiltonian adopted in Eq. (\ref{vs}) gives rise to
antisynchronization (that is, synchronization in antiphase).

For a quantitative characterization of the degree of
synchronization of the time dependent local spin observables,  we
will employ a time-correlation coefficient $C$, which can be
defined for any two time-dependent functions $f(t)$ and $f'(t)$ as
follows:
\begin{equation}
C_{f,f'}(t,\Delta t )=\frac{\overline{ \delta f \delta
f'}}{\sqrt{\overline{  \delta f^2} ~~\overline{\delta f'^2} }},
\label{sincroco}
\end{equation}
where the overbar stands for a time
average $\overline{f}=\int_{t}^{t+\Delta t}dt'f(t')/\Delta t$ with
time window $\Delta t$ and $\delta f=f-\overline{ f}$
{\cite{PRAsync}}. Phase-locked oscillations lead to
 $|C|= 1$, while it is easy to see that $C$ decreases to zero
for two signals displaying {uncorrelated} oscillations.
We shall take $\langle \sigma_1^x\rangle$ and $\langle
\sigma_2^x\rangle$ as local dynamical variables for the two spins,
and study the behavior of the synchronization
coefficient $C$ as a function of the
detuning $ \Delta=|\omega_2-\omega_1|$ and of the anisotropy
parameter $g$.

\begin{figure}[t]
\includegraphics[width=8cm]{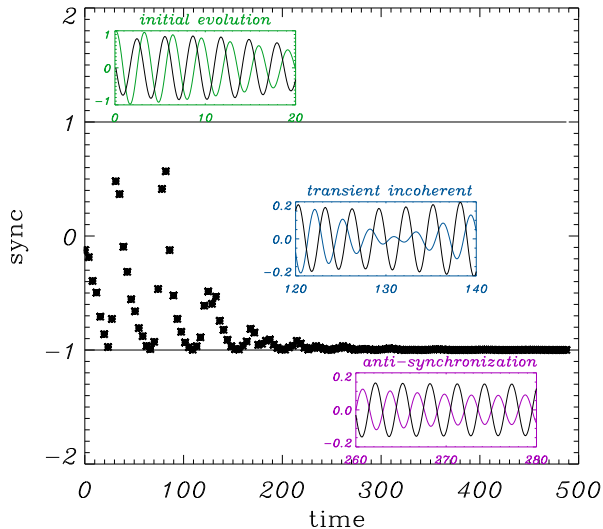}
\caption{(Color online) Synchronization coefficient $C_{\langle
\sigma_1^x\rangle,\langle \sigma_2^x\rangle}(t,\Delta t=6
\omega_1)$ (star symbols) as a function of time for $\omega_2=1.02
\omega_1$ and for $g=-1$. Synchronization is evaluated 
for partially overlapping time-average windows ($\Delta t=6
\omega_1)$,
at times $t=0,4\omega_1,\dots,500\omega_1$ [see Eq. (\ref{sincroco})].
 The bath temperature is $T=\omega_1$,
while the cut-off frequency is $\omega_c=20 \omega_1$. Finally,
the system-bath coupling strength is $\gamma=10^{-3} \omega_1$.
The initial state is $|\psi(0)\rangle=\left(\cos \theta_1 |
\uparrow\rangle + \sin\theta_1 | \downarrow\rangle\right)\otimes
\left(\cos \theta_2 | \uparrow\rangle + \sin\theta_2 e^{i \phi_2}|
\downarrow\rangle\right)$ with $\theta_{1}=\pi/4$,
$\theta_{2}=\pi/8$, and $\phi_{2}=\pi/2$ .
 Insets show the oscillatory evolution of
 $\langle\sigma_1^x(t)\rangle$ (lighter line)
and $\langle\sigma_2^x(t)\rangle$ (black line) for two different
times in the initial transient regime, and after
antisynchronization is reached. In all of the plot, time is in
units of $1/\omega_1$.
 \label{fig4}}
\end{figure}

An example of dynamically induced synchronization  is given in
Fig. \ref{fig4}. There, we consider the case in which the two
spins have similar frequencies ($\omega_2=1.02 \omega_1$) and
are initially prepared in an asymmetric factorized state and show
the behavior of $C_{\langle \sigma_1^x\rangle,\langle
\sigma_2^x\rangle}$ (star symbols) for a transversal coupling with
the bath ($g=-1$) as a function of time. In the early stage of the
evolution (upper inset in Fig. \ref{fig4}), the two functions  {
$\langle \sigma_1^x (t)\rangle$ and $\langle \sigma_2^x
(t)\rangle$}   oscillate with similar frequencies and
 with an increasingly different phase. Then, after a
transient incoherent time window (middle inset  in Fig. \ref{fig4}), finally
$C_{\langle \sigma_1^x\rangle,\langle \sigma_2^x\rangle}$
approaches $-1$ and the oscillations become antisynchronized
(details in the lower inset).

\begin{figure}[t]
\includegraphics[width=8cm]{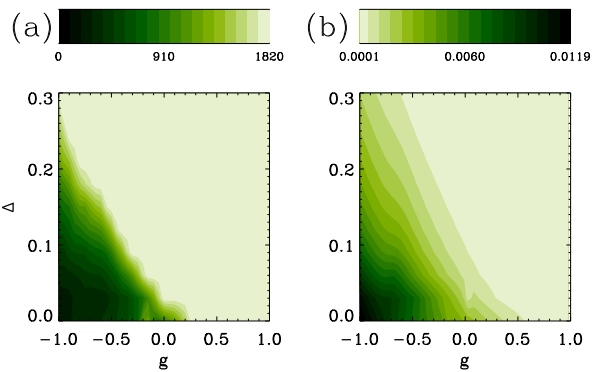}
\caption{(Color online) Synchronization maps vs. spins detuning
$\Delta$ (in units of $\omega_1$) and anisotropy coefficient $g$.
(a) Synchronization time $t_{synch}$, obtained by solution of Eq.
\ref{roab}. This is the time it takes for the synchronization
quality factor $C_{\langle \sigma_1^x(t)\rangle,\langle
\sigma_2^x(t)\rangle}$ to reach the threshold value $|C|=0.92$.
Darker colors correspond to synchronous dynamics emerging after a
short transient. (b) Difference between the two lowest (real parts
of the) eigenvalues of the Redfield tensor $R$
($\lambda^R_{(1)},\lambda^R_{(2)}$) with the higher weight in the
subspace spanned by $ \sigma_1^x$ and $ \sigma_2^x$ (see text for
details). The similarity between the two maps shows that
separation between damping rates in the system, appearing for $g <
0$, allows for the emergence of synchronization, which is found to
occur for dissipative coupling to the environment and small
detunings between the spins. No regime of synchronous oscillation
is found if dephasing dominates ($g>0$), irrespective of the
detuning. Bath temperature, cut-off frequency, system-bath
coupling,
 and initial state are the same as in Fig. \ref{fig4}.
\label{fig5}}  \end{figure}

A full characterization of the emergence of synchronization is
given in Fig. \ref{fig5}(a) where, by varying both  $ \Delta$ and
$g$, we calculate the time $t_{synch}$ after which $|C|$ reaches
the (arbitrarily fixed) threshold value $|C|=0.92$. As expected
from the comparison with the case of detuned harmonic oscillators
of Ref. {\cite{PRAsync}}, small values of the initial detuning
guarantee shorter synchronization times. If the two frequencies
are too different, the two spins are not able to synchronize
before reaching their steady state.
Furthermore, it is clear from this plot that if the environment coupling has a
more dephasing nature (that is, for $g>0$) synchronization does
not take place. In particular, for the purely dephasing case
discussed above [see Eq. (\ref{romn})], despite the presence of a
decoherence free subspace, it turns out that only {\it nonlocal
coherences} survive, while {\it local} spin observables always
decay irrespectively of the detuning and of the initial
conditions, thus showing that a substantial amount of relaxation
is essential for the emergence of synchronization.

\subsection{Synchronization and dynamical eigenvalues}\label{dyn}
According to Refs. \cite{PRAsync,3HO,sync_net},
the emergence of synchronization can be explained by considering
that, because of dissipation, after a transient time, only the
least-damped dynamical eigenmode survives. In order to find a
quantitative link between synchronization and the existence of
such a transient phase during which only one of the  eigenmodes is
active, we analyzed the behavior of the eigenvalues of the
Redfield tensor, represented as a $16\times 16$ matrix
($R_{abmn}\to R_{ab,mn}$), as a function of the detuning $\Delta$
and of the anisotropy $g$. Of the $16$ eigenvalues, $12$ are
complex (appearing in complex conjugate pairs) while four are real
numbers. These real eigenvalues only influence decays and do not
play any role as far as oscillatory dynamics is concerned; then,
we shall focus only on {the} complex ones.

All of the eigenmodes are damped, and the asymptotic dynamics is
governed by the least damped one, with the eigenvalue's real part
closest to zero.  A significant synchronization window, then,
should occur in the case in which the two (pairs of) eigenvalues
with largest real parts (smallest in absolute value), say
$\lambda_{(1)} = - \lambda_{(1)}^R \pm i \lambda_{(1)} ^I$ and
$\lambda_{(2)} = - \lambda_{(2)}^R \pm i \lambda_{(2)}^I$, satisfy
the relation $\lambda_{(2)}^R \gg \lambda_{(1)} ^R$.

In general, however, monitoring the two smallest real parts of
the eigenvalues of $R$ is not enough as, in many cases
the corresponding eigenmodes do not give a relevant
contribution to the expression for the local observables. An
explicit and noteworthy example of such a situation occurs for the
purely dephasing dynamics that is unable to give rise to
synchronization despite the presence of decoherence-free modes.

Then, we considered the two pairs of eigenmodes having more
influence on the specific observables we are interested in; and,
specifically, among the real parts of the three eigenvalues giving
rise to the slowest decays, we selected the two entering
$\sigma_1^x(t)$ and $ \sigma_2^x(t)$ with the highest weights. The
difference between the two decay rates selected in such a way,
$\lambda^R_{(1)}-\lambda^R_{(2)}$, is plotted in Fig.
\ref{fig5}(b) as a function of $g$ and $\Delta$. By comparing the
panels (a) and (b) in Fig. \ref{fig5}, we observe that there is a
very good qualitative agreement between $t_{synch}$ and
$\lambda^R_{(1)}-\lambda^R_{(2)}$, which confirms our
understanding of the emergence of synchronization as a result of
the presence of a slowly decaying mode giving a substantial
contribution to local spin components.

Since synchronization is due to the robustness of an  eigenmode of
the Bloch-Redfield tensor, the synchronization frequency
$\omega_{sync}$ of the two spins is expected  to be (very close
to) the imaginary part of the corresponding eigenvalue. The
agreement between the imaginary part of the eigenvalue
corresponding to the mode that decays more slowly and the
synchronization frequency
is very good also in the presence of
relatively strong detuning. Considering, for instance, the case of
$\omega_2=1.15\omega_1$, assuming $\omega_c=20\omega_1$,
$T=\omega_1$, and $\gamma =10^{-3}\omega_1$, we have ${\rm
Im}[\lambda^R_{(1)}]\simeq 2.306 \omega_1$, while a numerical
estimation performed over $50$ antisynchronized cycles gives
$\omega_{sync}/\omega_1=2.306 \pm 0.001$ (where the error comes
from time discretization).

\subsection{Discussion}
By analyzing the behavior of both $t_{synch}$ and
$\lambda^R_{(1)}-\lambda^R_{(2)}$ in Fig. \ref{fig5}, it becomes
clear that there is a qualitative difference between a
dephasing-dominated and a dissipative interaction with the
environment (roughly corresponding to positive and negative
values of $g$).
A fully analogous behavior will be found when discussing classical
and quantum correlations in Sec. \ref{sec2}. We observe that, when
dissipative effects are strong ($g<0$) the system is able to
synchronize in a time $t_{synch}$ that is shorter if $\Delta$ is
small  and longer for large detunings. {As a consequence}, if
$\Delta$ is too big (with respect to $\omega_1$), the system
reaches its steady state before synchronization can take place.

On the other hand, observing the right parts of Figs. \ref{fig5}(a)
and \ref{fig5}(b), we can conclude that if dephasing effects
prevail ($g\gtrsim 0$) synchronization does not take place,
independently of the detuning. For pure dephasing, in particular,
this can be seen directly by looking at the structure of  Eq.
(\ref{romn}), which implies a common decay factor for both {
$\langle\sigma_1^x(t)\rangle$ and $\langle\sigma_2^x(t)\rangle$}
and the persistence  of the two oscillation frequencies
$\omega_1$ and $\omega_2$ in the whole transient regime.

One could argue about the dependence of the synchronization
diagram of Fig. \ref{fig5}(a) on the initial state. Actually,
given the role played by dissipation, the  phenomenon is robust
against changes in the initial conditions.
Synchronous dynamics only arises
after the relaxation has washed out any sign of the initial state.

Another possible issue concerns the dependence of synchronization
on the bath temperature. While in the case of harmonic oscillators
described in Ref. {\cite{PRAsync}} the decay rates of the master
equation are temperature-independent, this is not true for the
spin master equation discussed in this paper. Given the connection
between decay rates and synchronization, temperature could be
imagined to play a role in determining synchronization times. We
analyzed this issue in detail by studying various thermal
regimes, but found no qualitative changes with respect to the
physical picture given above. In particular, an
increase in temperature (with respect to the case $T=\omega_1$ of
Fig. \ref{fig5}) would lead to an increase of $t_{synch}$, but
only for large values of the detuning. So, short synchronization
times would be almost unaffected, while long synchronization  times
would become longer by increasing the temperature of the bath.

\section{Quantum correlations}\label{sec2}
Having established the regimes in which synchronization takes
place in the open dynamics of our spin pair, we now explore a possible
connection between the emergence of synchronous
oscillations of the two {subsystems} and the appearance of
asymptotic quantum correlations between them. We will measure the
correlations using either entanglement or quantum discord (whose
definitions are briefly recalled below)
to show that their connection with synchronization is guaranteed by
dissipation being progressively lost when dephasing overcomes losses.

\subsection{Entanglement and discord}
Various indicators have been proposed for the degree of
quantumness of the state of a {bipartite} system. Among them, the
entanglement of formation ${\it E}_F$ is a well established
measure that quantifies the number of singlet states that are
necessary to prepare a given entangled state using only local
operations and classical communication \cite{horo}. For the case
of two qubits, it enjoys an analytic expression thanks to its
monogamic relationship with the concurrence \cite{wootters}. The
latter can be calculated  as ${{\it
E}_C=\max(0,\mu_1-\mu_2-\mu_3-\mu_4)}$, where $\mu_i$ are the
eigenvalues of the Hermitian matrix
$R=\sqrt{\sqrt{\rho}\tilde{\rho}\sqrt{\rho}}$ in decreasing order
($\tilde{\rho}$ is the spin flipped matrix of $\rho$). Finally,
${\it E}_F=H[(1-\sqrt{1-{\it E}_C^2})/2]$, where $H$ is the binary
entropy.

Recently, another {quantumness} quantifier, the quantum discord,
has attracted a lot of interest and attention due to its relevance
in quantum computing tasks \textit{not} relying on
entanglement~\cite{info-no-ent,modi}. Given a bipartite state
$\rho_{ab}$, its quantum discord $\delta_{a:b}$ is defined as the
difference between two inequivalent quantum versions of the
classical mutual information. Quantum mutual information, which is
assumed to capture the total amount of correlations between the
two parties $a$ and $b$ is defined as ${\cal I}(\rho_{ab})=
S(\rho_a)+S(\rho_b)-S(\rho_{ab})$, where $\rho_j$ is the
reduced density matrices of subsystem $j=a,b$ and $S(\rho_j)=-{\rm
Tr}\{\rho_j \log \rho_j\}$ is its von Neumann entropy.

According to Refs.~\cite{zurek,henderson}, ${\cal I}$ can be
divided into its classical part ${\cal C}_{a:b}$ and its quantum
part $\delta_{a:b}$. Classical correlations are given by ${\cal
C}_{a:b}(\rho)=\max_{\{E_i^b\}}\left[S(\rho_a)-S(a|\{E_i^b\})\right]$,
where the  conditional entropy is defined as
$S(a|\{E_j^b\})=\sum_ip_iS(\rho_{a|E_i^b})$, $p_i={\rm
Tr}_{ab}(E_i^b\rho)$, and where $\rho_{a|E_i^b}= {\rm Tr}_b
E_i^b\rho/{p_i} $ is the density matrix after a positive operator
valued measurement (POVM) $\{E_j^b\}$
has been performed on $b$.
Quantum discord is then defined as the difference between ${\cal
I}$ and ${\cal C}$:
$\delta_{a:b}(\rho)=\min_{\{E_i^b\}}\left[S(\rho_b)-S(\rho_{ab})+S(a|\{E_i^b\})\right]$.
Notice that both $\delta_{a:b}$ and ${\cal C}_{a:b}$ are
asymmetric under the exchange of the two parties, i.e., ${\cal
C}_{a:b}\neq{\cal C}_{b:a}$ and $\delta_{a:b}\neq\delta_{b:a}$.

In order to evaluate $\delta_{a:b}$, a minimization over all
possible POV's has to be performed. In the case of qubits,  the
optimal projective measurement  will  have between two and four
rank-$1$ elements \cite{dariano} (the case of two elements
corresponds to orthogonal measurements). Actually, as shown in
Ref. \cite{epl}, orthogonal measurements are sufficient for almost
all the states, and where three and four element POVMs outperform
them, the numerical difference  is always  very small and
negligible in qualitative analysis. Then, in the following, we
shall calculate the discord by limiting the minimization to
orthogonal projectors, as usually done in the literature.

\subsection{Long time behavior of entanglement}

\begin{figure}
\includegraphics[width=8cm]{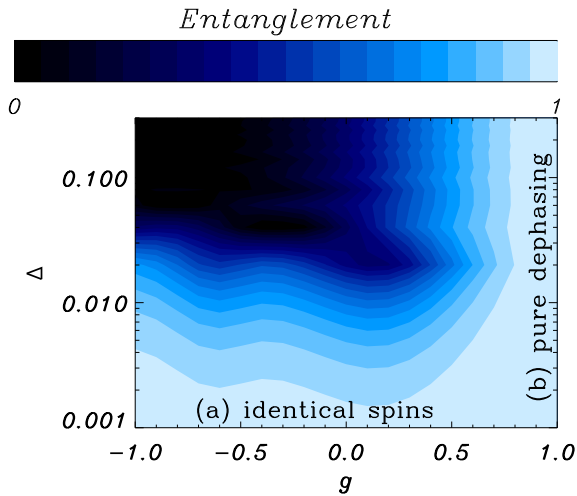}
\caption{(Color online)  Entanglement of formation at a time
$\omega_1 t=100$ for the case in which the initial state is the
singlet, $|\psi^-\rangle$. The robust character of
$|\psi^-\rangle$ against decoherence is shown both in the case of
identical spins and for pure dephasing, and is a direct
manifestation of the occurrence of a decoherence-free dynamics. On
the vertical axis, $\Delta$ is taken in units of the frequency
$\omega_1$.
Fast decoherence would be present with any parameter when
starting from most states orthogonal to the singlet, such as a factorized symmetric state.
As in the cases discussed before, the bath temperature is $T=\omega_1$,  the cut-off frequency is
$\omega_c=20 \omega_1$, while $\gamma=10^{-3}\omega_1$.
 \label{figadded}}
\end{figure}

Following the discussion in the previous sections, there are two
cases in which asymptotic entanglement is expected, provided the
initial state is not orthogonal to the singlet state. These are
the two cases in which the singlet is found to be a
decoherence-free state, namely, (a) for a detuning close to
zero, and (b) when dephasing prevails. As shown in Fig.
\ref{figadded},  this is indeed the case: for an initial singlet
state, an asymptotic entanglement is found both for $\Delta$ close
to zero and for $g$ close to unity.
Out of these specific parameter conditions, the system dynamics
{displays} a rather fast decoherence, leading to disappearance of
entanglement. As a robust entanglement is found to exist under
both of these conditions, it is clear that its presence has
nothing to do with synchronization. On the other hand, this is not
the case for more general quantum correlations.

\subsection{Discord dynamics}
Even when entanglement disappears from the system, quantum
correlations described by discord can be present and even
significantly large. In the following, we explore the temporal
dynamics of quantum correlations in different regimes and, in
particular, its dynamic generation starting from an initial
uncorrelated state. Two completely different scenarios appear in
the purely dephasing and dissipative limits, concerning the
buildup (and subsequent decay) of both classical correlations and
quantum discord, in particular, with respect to the dependence on
the detuning between the spins. The two regimes exactly correspond
to the emergence (or not) of spontaneous synchronization between
the spins.

\begin{figure}
\includegraphics[width=8cm]{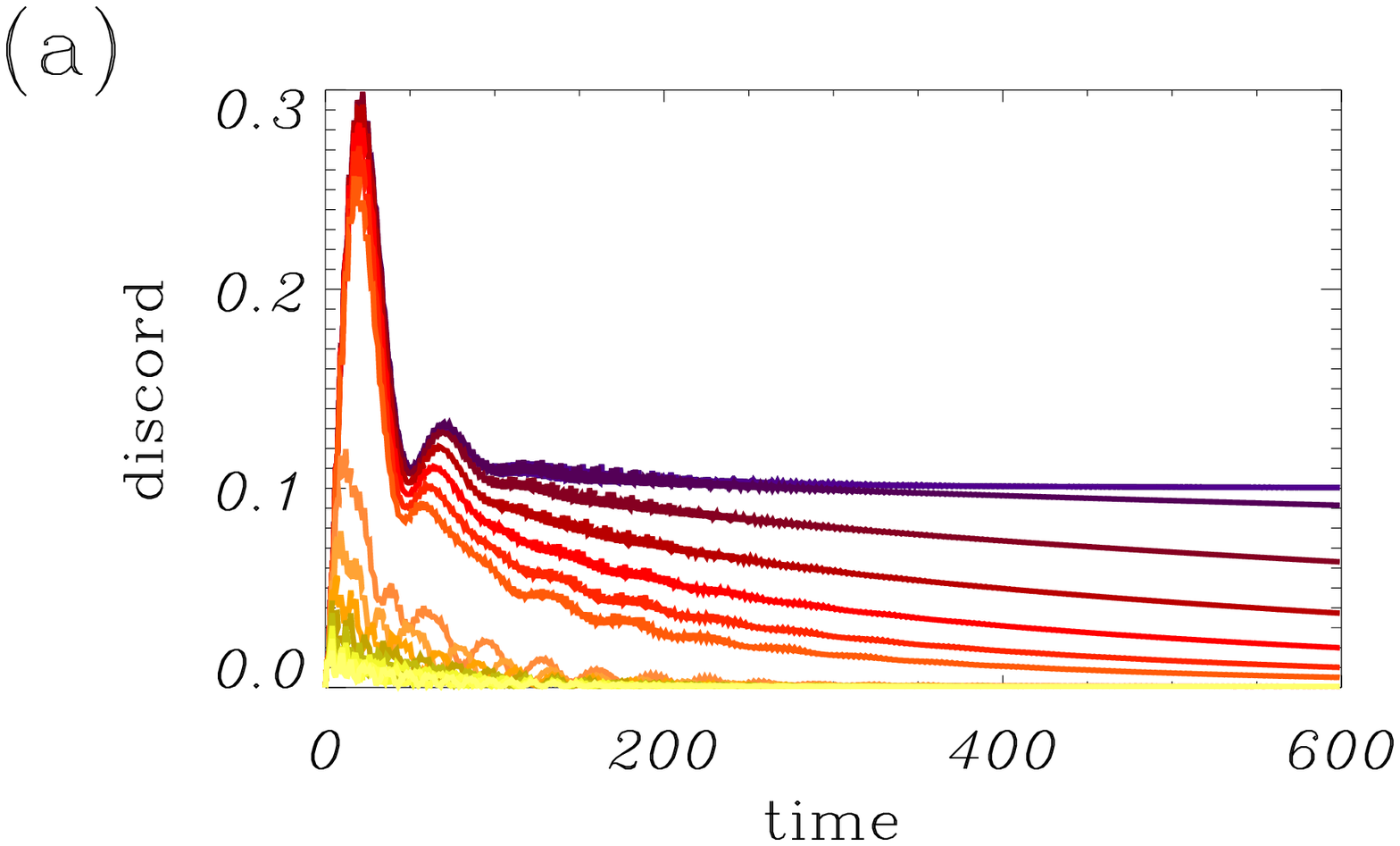}
\includegraphics[width=8cm]{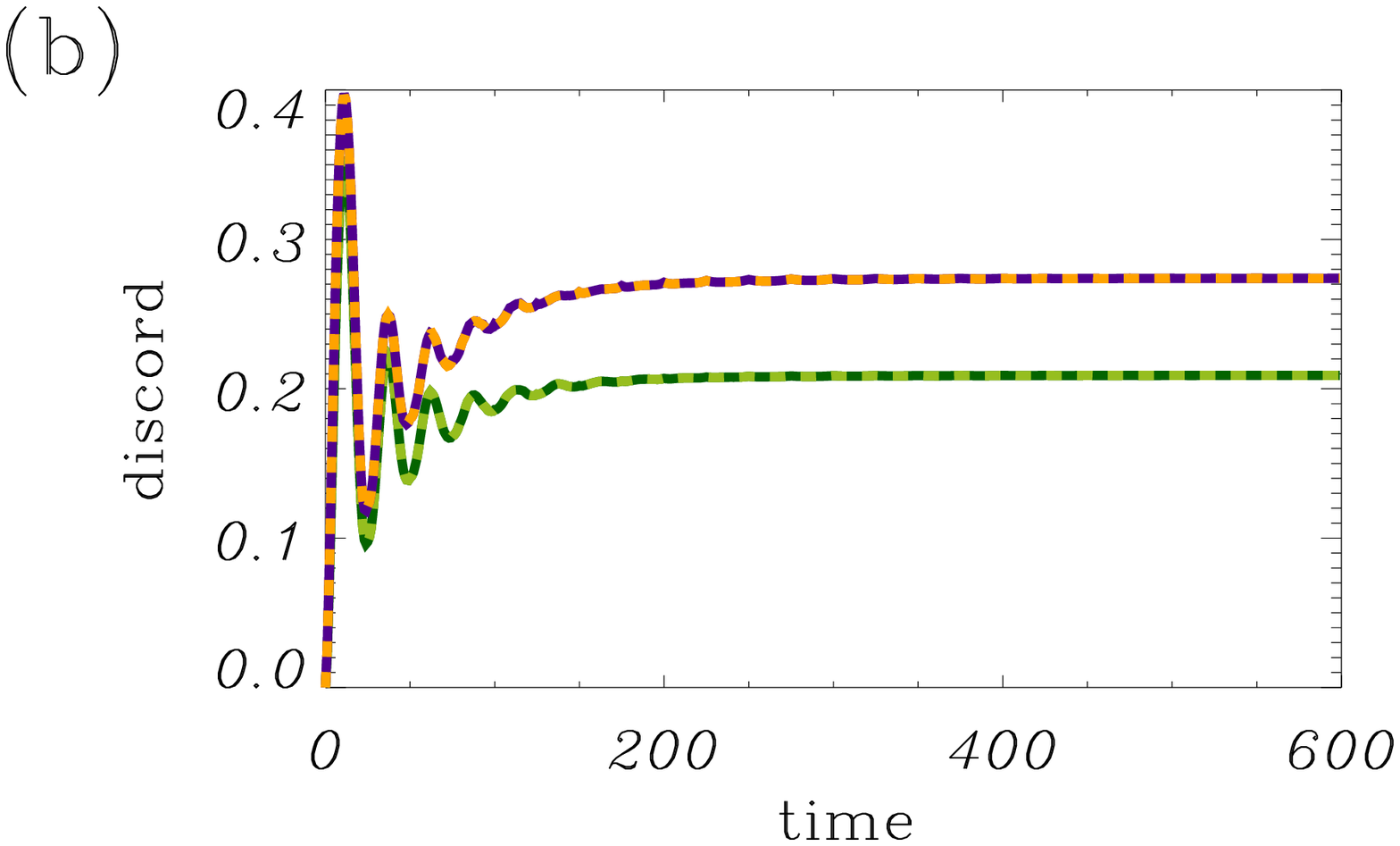}
\caption{(Color online) Dynamical generation of quantum discord
($\delta_{a:b}$) starting from a product state for different
detunings. (a) Dissipative bath with $g=-1$. Differently {colored}
lines correspond to different detunings $\Delta$ from a maximum of
$\Delta=1.25 \omega_1$ (lower curve in light color) to $\Delta=0$
(higher curve in dark color) with intermediate values
$\Delta/\omega_1 = \{n \cdot 0.005\}_{n=1,6}$ for the upper curves
and $\Delta/\omega_1= \{n \cdot 0.025\}_{n=1,9}$ for the lower
ones. The initial state is taken to be $|\psi(0)\rangle=(\cos
\theta_1 |\uparrow\rangle + \sin \theta_1
|\downarrow\rangle)\otimes (\cos \theta_2 |\uparrow\rangle + \sin
\theta_2 |\downarrow\rangle)$ with $\theta_1=\pi/3.2,~
\theta_2=\pi/3$. (b) Purely dephasing dynamics with $g=1$. The
time evolution of quantum discord is independent of the detuning
(different $\Delta$ give rise to superimposed curves). The two
lines correspond to different initial states, $\theta_1=\pi/3.2,~
\theta_2=\pi/3$ for the upper curve (which are the same values
used in panel (a), and $\theta_1=\pi/4 ,~ \theta_2=\pi/8$  for the
lower curve. In these plots, time is taken in units of
$\omega_1^{-1}$.  The bath temperature is $T=\omega_1$, the
cut-off frequency is $\omega_c=20 \omega_1$, and
$\gamma=10^{-3}\omega_1$. Time is expressed in units
$\omega_1^{-1}$.}\label{fig2}
\end{figure}

Indeed, as shown in Fig. \ref{fig2}, two distinct behaviors of
quantum discord are found, depending on the relative weight of the
transverse and longitudinal components of $V_S$ (inducing
relaxation and dephasing, respectively). Starting from a
nonsymmetric factorized state, and for negative values of the
anisotropy coefficient $g$, where the fully dissipative term
$\sigma_1^x+\sigma_2^x$ dominates, we observe that quantum
discord, apart from an abrupt initial increase, shows a monotonous
relaxation towards its equilibrium value. The decay of quantum
correlations is deeply influenced by the detuning. Indeed, for
nearly identical spins (that is, for $\omega_2/\omega_1$ close to
unity, corresponding to the synchronization region), relaxation is
characterized by a long transient regime where quantum discord has
a very small decay rate and remains almost frozen. The dynamical
evolution of $\delta_{a:b}$ for $g=-1$ and for various detunings
 in Fig.
\ref{fig2}(a) shows that the smaller the
detuning the higher the ``quasistationary" value of the discord
maintained during such a long-lasting transient regime.

\begin{figure}
\includegraphics[width=8cm]{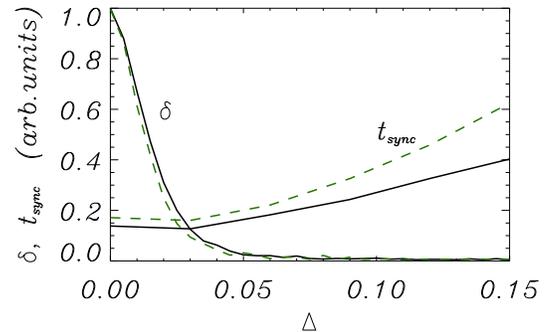}
\caption{(Color online) Value of quantum discord
[$\delta_{a:b}(t=300)$] and synchronization time $t_{synch}$
starting from a symmetric product state for $g=-1$ (continuous
lines) and $g=-0.8$ (dashed lines) when increasing the detuning
$\Delta$ (expressed in $\omega_1$ units). For the sake of
comparison, discord and synchronization time are rescaled. As
before, the bath temperature is $T=\omega_1$, the cut-off
frequency is $\omega_c=20 \omega_1$, and $\gamma=10^{-3}\omega_1$.
 \label{figdisco}}
\end{figure}

On the other hand, if the detuning $\Delta$ is too large compared
with the dissipation rates, the system is not able to build up
large enough correlations in the initial evolution and a quick
decay of quantum discord is observed. This is illustrated
comparing synchronization time with discord after an initial
transient in cases where dissipation prevails ($g=-1$ and $g=-0.8$
)  (Fig.\ref{figdisco}). We see that the larger the time taken by
the system to synchronize (worse synchronization) the smaller the
value maintained by discord. A similar behavior is found for the
classical correlations, which, however, are generally smaller than
the discord in this system. From this analysis, we conclude that
the establishment of transient quantum correlations and the
emergence of synchronization are strictly linked in this regime.

A completely different scenario emerges once positive values of
$g$ are taken into account. In Fig. \ref{fig2}(b) we consider the
other extreme case of pure dephasing ($g=1$). In this case, the
discord transient dynamics does not depend on $\Delta$, while its
asymptotic value changes for different initial conditions. As we
can see from Fig. \ref{fig2}(b), the dephasing channel is able to
generate an asymptotically robust amount of quantum discord
starting from a product state (the same is true for classical
correlations). On the other hand, it would be impossible for this
channel to build up entanglement.

The asymptotic value of discord can be
calculated by considering that the channel maps the initial state
$\rho(0)$ onto
\begin{equation}
\rho^{\infty}=\left(\begin{array}{cccc}
\rho_{11}(0) & 0 & 0 & 0\\
0 & \rho_{22}(0) & \rho_{23}(0)e^{-i \xi} & 0\\
0 & \rho_{32}(0)e^{i \xi} & \rho_{33}(0) & 0\\
0 & 0 & 0 & \rho_{44}(0)
\end{array}\right),
\end{equation}
where $\xi=2 \Delta t$, and with the density matrix written in the
computational basis $\{|\uparrow \uparrow\rangle, |\uparrow
\downarrow\rangle,|\downarrow \uparrow\rangle,|\downarrow
\downarrow\rangle \}$. An analytic expression for discord and
classical correlations of this class of states can be obtained by
using the results of Chen \textit{et al.} in Ref. \cite{chen}, who
showed that the conditional entropy is minimized either by using
the eigenstates of $\sigma_x$ or the ones of $\sigma_z$ to perform
the measurement on party $b$. In our case, the optimal measurement to
be performed in order to obtain the minimum conditional entropy is
given by the projections along the eigenstates of $\sigma_x$. The
maximum  achievable discord, for an initially factorized state, is
obtained if  $\rho_{ij}(0)=1/4$ for any of the non-empty entries
of $\rho^{\infty}$. Its value is $\delta_{\max}\simeq 0.312$ and
the corresponding classical correlations are ${\cal
C}_{\max}\simeq 0.188$.

As follows from our previous considerations, in the case of
identical spins the system will not completely thermalize for any
value of $g$, due to the presence of the decoherence-free singlet
state, unless it is initially prepared in a state orthogonal to
$|\psi^-\rangle$. With the exception of  this special case with
$\Delta=0$, and of the purely dephasing case $g=1$, the system
always reaches a thermal equilibrium state in a time that becomes
shorter and shorter as the detuning increases. For small
detunings, however, the transient regime displays a large time
window where robust quantum correlations are found, in spite of
dissipation.

\begin{figure}
\begin{center}
\includegraphics[width=8cm]{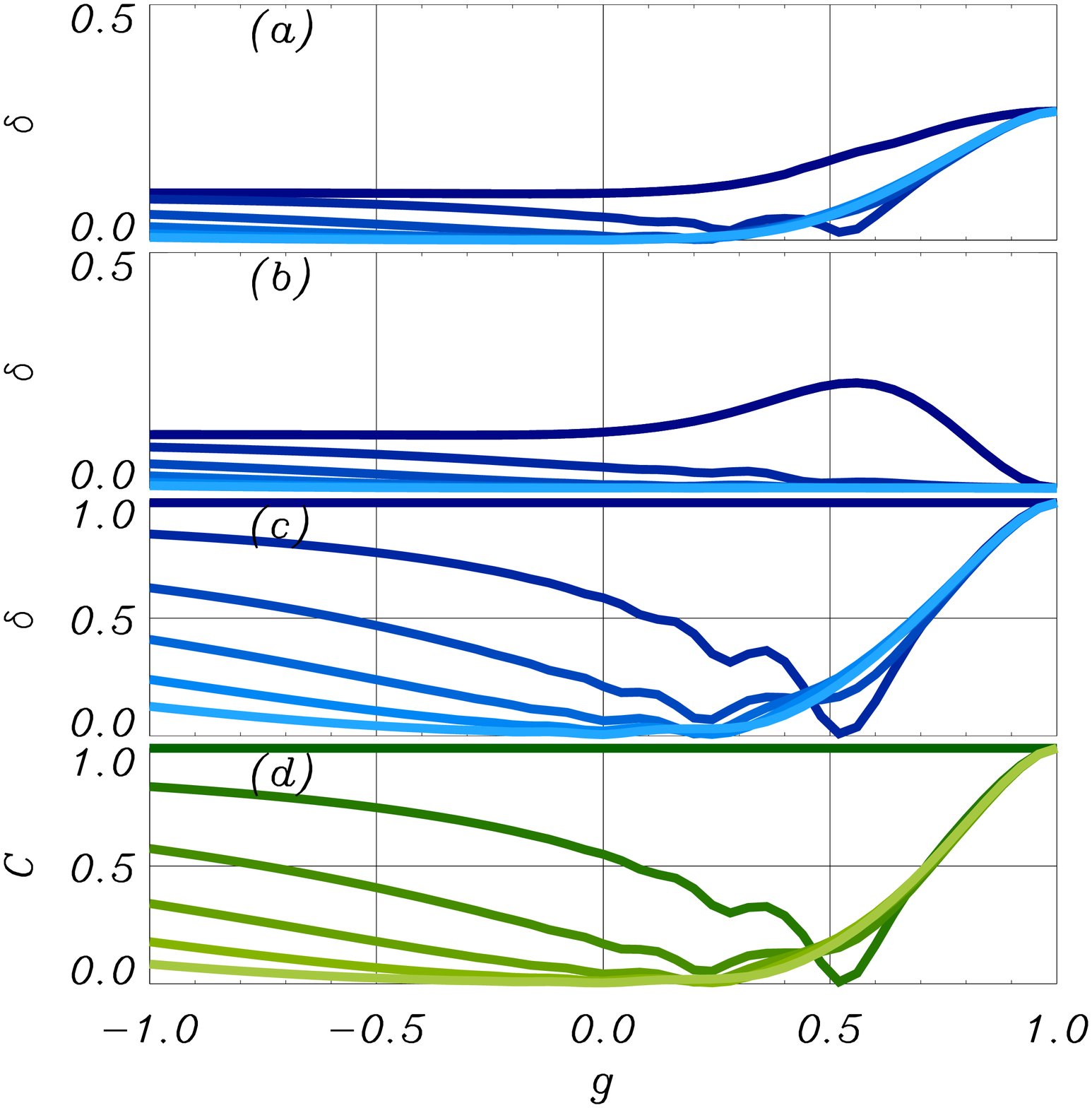}
\caption{(Color online)  Long-time behavior of discord $\delta$
and of classical correlations ${\cal C}$, evaluated at
$t=800/\omega_1$ as a function of the anisotropy parameter $g$. In
all of the panels, the six different curves correspond to
detunings increasing from $\Delta=0$ (higher curves) up to
$\Delta=0.025 \omega_1$ (lower curves). The different panels
correspond to different initial conditions: in (a) we have chosen
a product state with $\theta_1=\pi/4$, $\theta_2=\pi/8$; in (b)
the maximally entangled state {$|\phi^+\rangle$}, while panels (c)
and (d) are obtained by taking {$|\psi^-\rangle$} as the initial
state of the spin pair. The bath temperature and the cut-off
frequency are the same as in any other plot ($T=\omega_1$  and
$\omega_c=20 \omega_1$), while $\gamma=10^{-3}\omega_1$.
\label{fig3}}
\end{center}
\end{figure}

The long-time behavior of quantum discord and of classical
correlations for any value of $g$ is illustrated in Fig.
\ref{fig3}. Both $\delta$ and ${\cal C}$ are calculated, for the
fixed time $\omega_1 t=800$, as a function of $g$ and  for a set
of values of the detuning $\Delta$. The separation between the
``dissipative regime'' and the ``dephasing regime'' is clear from
this plot. In the ``dephasing regime" there is a coalescence of
all the lines and the detuning does not play any special role,
consistent with what is shown in Fig. \ref{fig2}(b). In contrast,
the qualitative behavior described for $g=-1$  in  Fig.
\ref{fig2}(a) persists up to around $g=0$. In other words, both
the discord and the classical correlations ${\cal C}$ display an
increasing robustness against dissipation as $\Delta$ decreases.
This behavior (with the dependence on $\Delta$ when dissipation
prevails and the robustness of correlations for small detunings)
is definitely analogous to what we have found for the
time-correlation coefficient describing synchronization.

\section{Conclusions}
We have investigated the long-time dynamics of two spins
interacting through a common thermal bath and have shown that,
depending on the relative weights of the environment induced
dissipation and dephasing, two qualitatively different dynamic
regimes are observed for both spin-spin correlations and mutual
synchronization.

The presence of dissipation induces a time scale separation in the
decay rates of the eigenmodes of the Redfield tensor, which govern
the system's evolution. This allows one to observe spontaneous
synchronization between the local observables of the two spins.
When the precessions of the two spins  are synchronous, long-time
classical and quantum correlations (as measured by quantum
discord) are found for the spin pair, which become more and more
robust against decoherence as the synchronization time gets
shorter (which is the case for small detunings).

On the other hand, a channel in which dephasing prevails is not
able to generate any time scale separation and cannot support any
kind of dynamical synchronization. Long-time (and even asymptotic)
quantum correlations between the spins may exist in this case, but
they have a completely different origin. Indeed, they are due to
the existence of a decoherence-free subspace which prevents the
decay of part of the initial coherences of the total system, while
allowing for the dynamic cancellation of some others, in a way
that enables the possibility of generating quantum correlations
even from an initially factorized state. Remarkably, this
mechanism does not allow for the generation of entanglement, but
only of quantum discord \cite{gener}.
When the dissipative nature of the coupling with the environment
prevails, instead, quantum discord is generated precisely because
of the emergence of synchronization.

\acknowledgments
This work was funded by MICINN, MINECO and FEDER under Grants
No. FIS2007-60327 (FISICOS) and No. FIS2011-23526 (TIQS), and by
Balearic Islands Government. G. L. G. acknowledges financial support by
Compagnia di San Paolo,
by Science Foundation of Ireland under Project No. 10/IN.1/I2979,
and by EU commission (IP AQUTE and STREP PICC).
The visiting professors program of the University of Balearic 
Islands and COST Action MP1209 are also acknowledged.


\begin{thebibliography}{0}
\bibitem{nota} The fact that the state $|\psi^-\rangle$ is free from
decoherence depends on the relative sign (or, in general, relative
phase) between $\sigma_1^x$ and $\sigma_2^x$ in the coupling term
in Eq. (\ref{vs}). It is immediately seen that, by applying
symmetry considerations, it would be $|\psi^+\rangle$ not to
decohere for a coupling of the form $\sigma_1^x-\sigma_2^x$.
\end{thebibliography}
\end{document}